\documentclass[draft,onecolumn]{IEEEtran}
\usepackage{setspace}
\usepackage{stackrel}
\usepackage[colorlinks,linkcolor=black,citecolor=black,urlcolor=black,bookmarks=false]{hyperref}
\usepackage[spanish,british]{babel}
\usepackage[utf8]{inputenc}

\usepackage{amsmath,amsthm,amsfonts,amssymb}
\usepackage[mathscr]{euscript}
\usepackage{graphicx,xcolor}
\usepackage{multicol,enumerate,fancyhdr,emptypage,xspace,subfig}

\usepackage{indentfirst}
\usepackage{multirow}
\usepackage{mathtools}

\DeclareMathOperator{\WF}{WF}

\DeclareMathOperator{\supp}{supp}
\DeclareMathOperator{\wt}{\mathrm{wt}}
\DeclareMathOperator{\im}{\mbox{\rm{Im}}}

\newcommand{\Z}{{\mathbb Z}}
\newcommand{\FFF}{{\mathbb F}}

\newcommand{\CCC}{{\mathcal C}}
\newcommand{\ie}{\emph{i.e.}}

\newcommand{\dd}{\rotatebox{0}{\scalebox{0.75}{$\ddots$}}}

\newcommand{\vect}[1]{\mathbf{#1}}
\newcommand{\vv}{\vect{v}}
\newcommand{\uu}{\vect{u}}
\newcommand{\ee}{\vect{e}}

\newcommand{\yy}{\vect{y}}

\newenvironment{sbmatrix}[1]{\left[\begin{array}{#1}}{\end{array}\right]}

\newcommand{\DD}{(D^{-1},D)} 
\newcommand{\DDD}{(D)} 
\newcommand{\DDDD}{[D^{-1},D]}

\newtheorem{theorem}{Theorem}
\newtheorem{theorem*}{Theorem*}

\newtheorem{example}[theorem]{Example}
\newtheorem{lemma}[theorem]{Lemma}
\newtheorem{notation}[theorem]{Notation}

\newtheorem{remark}[theorem]{Remark}
\newtheorem{Properties}[theorem*]{Properties}

\begin{document}
%
\title{Smaller Keys for the McEliece Cryptosystem: A Convolutional Variant with GRS Codes}
	\author{P. Almeida, M. Beltr\'{a}, D. Napp and C. Sebasti\~{a}o}

\thanks{ Paulo Almeida and Cl\'{a}udia Sebasti\~[{a}o,
	CIDMA - Center for Research and Development in Mathematics and Applications, Dept.\ of Mathematics, University of Aveiro, Portugal, palmeida@ua.pt and claudia.sebastiao@ua.pt\\
	\\
	This paper was presented in part at the The 23rd International Symposium on Mathematical Theory of Networks and Systems (MTNS 2018), Hong Kong.\\
	\\
	Miguel Beltr\'{a} and Diego Napp,  Dept.\ of Mathematics, University of  Alicante, Spain,  miguel.beltra@ua.es and diego.napp@ua.es
}

%



\maketitle

\begin{abstract}
 	
    In this paper we present a variant of the McEliece cryptosystem that possesses several interesting properties, including a reduction of the public key for a given security level. In contrast to the classical McEliece cryptosystems, where block codes are used, we propose the use of a convolutional encoder to be part of the public key. The permutation matrix is substituted by a polynomial matrix whose coefficient matrices have columns with weight zero or at least weight two. This allows the use of Generalized Reed-Solomon (GRS) codes which translates  into shorter keys for a given security level. Hence, the private key is constituted by a generator matrix of a GRS code and two polynomial matrices containing large parts generated completely at random. In this setting the message is a sequence of messages instead of a single block message and the errors are added throughout the sequence. We discuss possible structural and ISD attacks to this scheme. We conclude presenting the key sizes obtained  for different  parameters and estimating the computational cost of encryption and decryption process.
 	
\end{abstract}

\begin{IEEEkeywords}
Convolutional Codes, Generalized Reed-Solomon Codes, Code-based Cryptography,  McEliece Cryptosystem, Information Set Decoding.
\end{IEEEkeywords}

\section{Introduction}

Code-based public-key cryptosystems (PKC) are considered promising alternatives for public-key cryptography since their security relies on well-known NP-hard problems and allow fast encryption and decryption procedures. Moreover, no feasible attacks to these cryptosystems based on quantum algorithms are known, unlike the schemes based on integer factorization or the computation of the discrete logarithm  for which attacks based on the Shor's algorithm could be successfully performed \cite{Shor}, and therefore are candidates for post-quantum cryptography. However, one of the main disadvantages of code-based schemes is the large keys whose size is inherently determined by the underlying Goppa block code used in the original cryptosystem.

\medskip

For this reason, there have been several attempts to substitute Goppa codes by other classes of block codes, \emph{e.g.}, Generalized Reed-Solomon (GRS), Low-Density Parity-Check, Quasi-Cyclic, Algebraic Geometry codes among others \cite{Baldi2018,BBCRS,Berger2005,BLP08,GR2017,Pellikaan2017} and \cite{Tillich_IEEE2013}. GRS codes are Maximum Distance Separable (MDS) codes \cite[Ch.10, \S 8]{ma77} which, in the McEliece scheme, translates into smaller key sizes, see \cite{BBCRS} and \cite{nied86}. Hence, a major improvement would be achieved if these codes could be securely used in these cryptosystems. Unfortunately, due to their strong algebraic structure, they are vulnerable to many structural attacks. One recent interesting idea to remove this algebraic structure was to replace the permutation used in the original McEliece cryptosystem with more general transformations, in  \cite{BBCRS,GR2017,MaTi2016,karan2018,karan2019,wang2016} and \cite{wang2017}. However, some of these variants have been partially or fully broken using attacks based on the (twisted) Schur square code distinguisher that permit to distinguish GRS codes from random ones in \cite{Tillich2015,Tillich_IEEE2017} and \cite{couvreur2020security}. 

\medskip

Another interesting variant was proposed in \cite{LoJo12} where the secret code is a convolutional code. 
One of the most appealing features of this system is the fact that its secret generator matrix contains large parts that are generated completely at random. 
But the authors already pointed out that this approach suffered from two main problems. The first one being the lack of efficient decoding algorithms, namely, that if one wants a maximum likelihood decoding, such as the Viterbi decoding algorithm, the memory used must be very limited. The second problem stems from the fact that convolutional codes usually start from the all zero state, and therefore the first code symbols that are generated will have low weight parity-checks, which would imply security threats. In fact, the scheme in \cite{LoJo12} had low weight codewords that revealed the underlying code structure and was broken in \cite{Tillich2013}.

\medskip

Hence, GRS codes are good candidates for McEliece PKC because of their small key sizes, but bad candidates because their pervasive structural properties are a security weakness. Convolutional codes are good candidates for McEliece PKC because of the randomness in their construction, but bad candidates because early transmissions will be low weight and exploitable. The aim of this manuscript is to combine the strengths of both in a way that addresses the security flaws of each. In particular, we present a new variant that allows to consider GRS codes using a convolutional mask. In our scheme, the plaintext is not a block vector but a stream of smaller vectors sent in a sequential fashion. The public key is given by a polynomial convolutional encoder $G'(D)=S(D)GP(D)$ where $G$ is the generator matrix of a GRS code, $S(D)$ a polynomial matrix and $P(D)$ a non singular polynomial matrix selected to protect against structural attacks providing weight two masking at every instant. As pointed out in \cite{GR2017} and \cite{karan2018} (see also \cite{karan2019}), the weight two masking seems to be enough for the removal of any identifiable algebraic structure from the public code. Hence, the proposed scheme  increases the security level against known structural attacks for block codes, and in particular against any distinguisher attack based on the Schur product. We note that the construction of $S(D)$ and $P(D)$ uses matrices with large parts generated completely at random and can be easily constructed.  Moreover, it seems that no obvious subcodes of the public code with small support can be computed in this proposal. We present several examples for comparison with previous variants of the McEliece cryptosystem to illustrate the key size reduction of the public key achieved by this novel scheme. As a main drawback of this approach, ciphertexts for this scheme are considerably longer than the ones used in most common public-key encryption schemes.


\section{Preliminaries}

This section contains the background needed for the development of our results. We introduce the basic definitions of convolutional codes and then recall the classical McEliece cryptosystem.

\medskip

Let $\FFF = \FFF_q $ be a finite field of size $q$, $\mathbb F[D]$ the ring of polynomials, $\FFF [D^{-1},D]$ the ring of Laurent polynomials and $\FFF(D)$ the field of rational polynomials, all with coefficients in $\mathbb F$. Notice that $\mathbb F[D]\subseteq \FFF\DDDD\subseteq \FFF(D)$. 
The \emph{support} of a vector $\vv =(v_1,\dots , v_n) \in \FFF^n$ is the set $\supp(\vv) = \{ i \in \{1,\ldots,n\} : v_i \neq 0 \}$,  and its size is called the \emph{Hamming weight} of $\vv$, denoted by $\wt(\vv)$. This definition can be extended to (rational) polynomial vectors $\vv\DD=\sum\limits_{i\in \Z} \vv_i D^i \in \FFF(D)$ in a natural way as $\wt (\vv\DD)=\sum\limits_{i\in \Z} \wt(\vv_i)$.



\medskip
\begin{notation}\label{js}
Given an integer $j$ and a positive integer $s$, we denote by $[j]_s$ the canonical representative of the residual class of $j$ modulo $s$, \ie, the smallest non negative integer that is congruent with $j$ modulo $s$.
\end{notation}

\begin{remark}
Throughout the paper, we will use the above notation for the indices of some vectors, so it is worthwhile to point out that, for $-s \leq j \leq 2s-1$,
\[
[j]_s =
\left\lbrace
\begin{array}{cl}
	j+s & \text{ when } -s \leq j \leq -1; \\
	j & \text{ when } 0\leq j\leq s-1; \\
	j-s & \text{ when } s\leq j\leq 2s-1.
\end{array}
\right.
\]
\end{remark}

\subsection{Convolutional codes}

As opposed to block codes, convolutional codes process a continuous sequence of data instead of blocks of fixed length vectors. If we introduce a variable $D$, usually called the \emph{delay operator}, to indicate the time instant, then we can represent the sequence message $(\uu_{0},\uu_{1}, \ldots,\uu_{s-1})$, $\uu_i \in \FFF^k$ as a polynomial  $\uu\DDD = \uu_{0}+\uu_{1} D +\uu_{2}D^2 + \cdots +\uu_{s-1}  D^{s-1} \in \FFF^k[D]$. In this representation the encoding process of convolutional codes, and therefore the notions of convolutional code and convolutional encoder, can be presented as follows. Note that unlike linear block codes, there exist several possible ways to define convolutional codes (see, for instance, \cite{LiPiRo,gl03,ro99a} for more details).

\medskip

A \textit{convolutional code} ${\mathcal C}$ of rate $k/n$ is an $\FFF[D]$-submodule of $\FFF^n[D]$ of rank $k$ given by a polynomial \emph{encoder matrix} $G(D) \in \FFF^{k \times n}[D]$,
\[
\mathcal{C} = \im_{\FFF[D]} G(D) = \{ \uu\DDD G(D) : \uu\DDD \in \FFF^k[D] \},
\]
where $\uu\DDD $ is called the {\em information vector}. If
\[
G(D) = \sum_{i=0}^m G_i D^i \in \FFF^{k \times n}[D]
\]
with $G_m\neq 0$, $m$ is called the \emph{memory} of $G(D)$, since it needs to ``remember" the inputs $\uu_i$ from $m$ units in the past. Note that when $m=0$ the encoder is constant, so it can be seen as a generator matrix of block code. Hence, the class of  convolutional codes generalizes the class of linear block codes in a natural way. Within this framework, a dual description of a convolutional code ${\mathcal C}$ in terms of a \textit{parity-check} matrix is not always possible \cite{ro96a1}. It exists only if $G(D)$ is \emph{basic}, \ie, has a polynomial right inverse.

\medskip

Let  $\uu(D) = \sum_{i=0}^{s-1} \uu_i D^i$ be an information vector and $G(D)$ an encoder of a convolutional code. Assume that the codeword $\uu(D)G(D)$ has been transmitted and  $\yy(D) = \sum_{i=0}^{s+m-1} \yy_i D^i$ has been received. If $\yy(D) \neq \uu(D)G(D)$ then some errors have occurred during the transmission. Let $\ee(D) = \yy(D) - \uu(D)G(D) = \sum_{i=0}^{s+m-1} \ee_i D^i$ be the polynomial vector representing these errors. 

\medskip

Sometimes it is useful to use matrix notation instead of polynomial notation. With this notation we can represent this encoding scheme as follows. Let
\begin{align*}
	\yy_{\rm total} & = \begin{sbmatrix}{cccccc}
		\yy_0 & \yy_1 & \cdots & \yy_{s+m-1}
	\end{sbmatrix} \in\FFF^{(s+m)n}, \\
	\uu_{\rm total} & = \begin{sbmatrix}{cccc}
		\uu_0 & \uu_1 & \cdots & \uu_{s-1}
	\end{sbmatrix} \in\FFF^{sk}, \nonumber \\
	\ee_{\rm total} & = \begin{sbmatrix}{cccc}
		\ee_0 & \ee_1 & \cdots & \ee_{s+m-1}
	\end{sbmatrix} \in\FFF^{(s+m)n}, \nonumber
\end{align*}
and
\begin{equation*}
	G_{\rm total} =
	\begin{sbmatrix}{cccccccccc}
		G_0 & G_1 & G_2 & \cdots & \cdots & G_m & & & & \\
		& G_0 & G_1 & G_2 & \cdots & \cdots & G_m &  & &  \\
		& & \ddots & \ddots & \ddots & & & \ddots  \\
		& & &  \ddots & \ddots & \ddots & & & \ddots  \\
		& & & & G_0 & G_1 & G_2 & \cdots &   \cdots & G_m
	\end{sbmatrix} \in\FFF^{sk\times(s+m)n},
\end{equation*}
Then, we have that $\yy_{\rm total} = \uu_{\rm total} G_{\rm total} + \ee_{\rm total}$. 

\subsection{The original McEliece cryptosystem}

Next we briefly recall the original McEliece PKC (see \cite{McEliece1978}). In the classical McEliece cryptosystem, Bob's private key is composed by an encoder $G\in \FFF^{k \times n}$ of an $(n,k)$ block code $\CCC$ capable of correcting $t$ errors, a non singular matrix $S \in \FFF^{k \times k}$ and a permutation matrix $P\in \FFF^{n \times n}$. The public key is $G' = SGP$. 
To encrypt a cleartext message $\textbf{u} \in \FFF^k$ Alice chooses a random error $\ee \in \FFF^n$ of weight $\wt (\ee)\leq t$ and computes the ciphertext
$$
\textbf{y} = \textbf{u}G' + \textbf{e} = \textbf{u}SGP + \textbf{e} .
$$
When Bob receives the vector $\textbf{y}$, multiplies it from the right by the inverse of $P$ and recovers $\textbf{u}S$ by decoding $(\textbf{u}S) G + \textbf{e} P^{-1}$, as $\wt(\textbf{e} P^{-1})\leq t$. Finally, Bob multiplies $\textbf{u}S$ by the matrix $S^{-1}$ to obtain $\textbf{u}$.

\section{A new variant of the McEliece PKC based on convolutional codes}\label{ConstrG}

We divide this section in three parts. In the first one, we present the general description of the cryptosystem: public and private keys, and encryption and decryption procedures. In the second part it is shown that the decryption procedure works. In the third part we instantiate the general scheme and consider $G$ as the generator matrix of a GRS code. For this class of codes we present a large family of matrices $S(D)$ and $P(D)$ satisfying the requirements of the proposed cryptosystem. The symbol $[j]_s$ defined in Notation \ref{js} will be used several times below.

\subsection{Keys, encryption and decryption}\label{subsection2.1}

\noindent \textbf{Public parameters}: The message length $s\geq 5$ and the ratio $t/\rho$ (see the description below about the private key for the definition of $t$ and $\rho$).

\medskip

\noindent \textbf{Private key}: It is composed by three matrices \{$G$, $S(D)$, $T\DD$\}:

\begin{itemize}
\item[{\bf 1)}] $G \in \FFF^{k \times n}$ is a generator matrix of an $(n,k)$ linear code having an efficient error-correction algorithm able to correct up to $t$ errors.

\item[{\bf 2)}] $S(D) \in \FFF^{k \times k}[D]$ is a polynomial matrix of the form $S(D) = S_0 + S_1 D + S_2 D^2$ such that the matrix $S_{\rm trunc}$ defined as
\[
S_{\rm trunc} =
\scalebox{0.8}{$
	\left[
	\begin{array}{cccccccccccccccccc}
		S_0 & S_1 & S_2 &&&&& \\
		& S_0 & S_1 & S_2 &&& \\
		&& S_0 & S_1 & S_2 && \\
		&&& \dd & \dd & \dd & \\
		&&&& \dd & \dd & S_2 \\
		&&&&& \dd & S_1 & S_2 \\
		S_2 &&&&&& S_0 & S_1 \\
		S_1 & S_2 &&&&&& S_0 \\
	\end{array}
	\right]
	$} \in \FFF^{sk \times sk},
\]
is non singular.

\item[{\bf 3)}] $T\DD  \in \FFF^{n \times n}[D^{-1},D]$ is an invertible Laurent polynomial matrix of the form
\begin{equation*}
	T\DD = T_{-2} D^{-2} + T_{-1} D^{-1} + T_0
\end{equation*}
such that each row of each coefficient matrix $T_i$, $i \in \{-2,-1,0\}$, has no more than $\rho$ nonzero elements and whose inverse is denoted by $P(D) = T^{-1}\DD$ and is of the form
\begin{equation*}
	P(D) = P_0 + P_1 D + P_2 D^2.
\end{equation*}
\end{itemize}

\medskip

\noindent \textbf{Public key}: A polynomial matrix $G'(D) \in \FFF^{k \times n}[D]$ defined as
\[ G'(D) = S(D) G P(D) = G'_0 + G'_1D + G'_2D^2 + G'_3D^3 + G'_4 D^4. \]

\medskip

\noindent \textbf{Encryption:}  To encrypt a message $\uu(D)= \uu_0 + \uu_1 D + \uu_2 D^2 + \cdots +\uu_{s-1} D^{s-1} \in \FFF^k[D]$, Alice selects at random an error vector $\ee(D) = \ee_0 + \ee_1 D + \cdots + \ee_{s-1} D^{s-1} \in \FFF^n[D]$ satisfying the condition
\begin{align}
\label{eq:weight_e}
&\wt([\ee_ {[i]_s} \  \ee_{[i+1]_s} \ \ee_{[i+2]_s}]) \leq \frac{t}{\rho}, && \hspace{-2.7cm} \text{ for all } i \in \{ 0,1,\ldots,s-1 \}.
\end{align}
Then Alice computes and sends the vector
\begin{align*}
	\yy(D) & = \uu(D) G'(D) + \ee(D) \pmod {D^s-1}.
\end{align*}

\medskip

\noindent \textbf{Decryption:} To decrypt a ciphertext $\yy(D)$, Bob computes the vector
\begin{align*}
	\hat{\yy}(D) & = \yy(D)T\DD \pmod{D^s-1}.
\end{align*}
The vector $\hat{\yy}(D)$ is of the form 	$\hat{\yy}_0 + \hat{\yy}_1 D + \cdots + \hat{\yy}_{s-1} D^{s-1}$ and each coefficient vector can be computed as
\[
\hat{\yy}_{i} = \yy_{[i]_s} T_0 + \yy_{[i+1]_s} T_{-1} + \yy_{[i+2]_s} T_{-2}, \hspace{0.5cm} i \in \{ 0,1,\ldots,s-1 \}. \]

It is shown in the next section that each $\hat{\yy}_i$ is of the form $\hat{\yy}_i = \hat{\uu}_i G + \hat{\ee}_i$, for some vectors $\hat{\uu}_i \in \mathbb{F}^k$, and $\hat{\ee}_i \in \mathbb{F}^n$ with $\wt(\hat{\ee}_i) \leq t$. Hence, Bob can recover each $\hat{\uu}_i$ using an efficient decoding algorithm of the linear code generated by $G$. Finally, it holds that
\begin{equation*}
\left[
\begin{array}{cccc}
    \hat{\uu}_0 & \hat{\uu}_1 & \cdots & \hat{\uu}_{s-1}
\end{array}
\right]
=
\left[
\begin{array}{cccc}
    \uu_0 & \uu_1 & \cdots & \uu_{s-1}
\end{array}
\right]
S_{\rm trunc},
\end{equation*}
and the message $\uu(D)$ can be recovered since $S_{\rm trunc}$ is non singular. 



\subsection{Details on the decryption process}\label{sec:decryption}

In this section we show why the decryption process explained above works. Condition \eqref{eq:weight_e} describes the maximum number of errors allowed within a time interval. We start with the following result that will be used later. 

\begin{lemma}\label{lem:01}
Let $T\DD$ and $\ee (D)$ as defined above. For each $i \in \{ 0,1,\ldots,s-1 \}$ consider the vector
\[
\hat{\ee}_{i} = \ee_{[i]_s} T_0 + \ee_{[i+1]_s} T_{-1} + \ee_{[i+2]_s} T_{-2}.
\]
Then $\wt(\hat{\ee}_i) \leq t$ for each $i \in \{ 0,1,\ldots,s-1 \}$.
\end{lemma}

\begin{proof} Each row of $T_j$, $j \in \{-2,-1,0\}$, has at most $\rho$ nonzero elements so $\wt (\ee_r T_j) \leq \rho \cdot \wt (\ee_r)$ for all $r \in \{ 0,1,\ldots,s-1 \}$. Since $\hat{\ee}_{i} = \ee_{[i]_s} T_0 + \ee_{[i+1]_s} T_{-1} + \ee_{[i+2]_s} T_{-2}$  then \eqref{eq:weight_e} implies $\wt(\hat{\ee}_i) \leq t$.
\end{proof}

Using matrix representations it is easy to check that the ciphertext can be decomposed as follows:
\begin{equation}\label{eq:ciphertext_trunc}
	\left[
	\begin{array}{cccc}
		\yy_0 & \yy_{1} & \cdots & \yy_{s-1}
	\end{array}
	\right]
	=
	\left[
	\begin{array}{cccc}
		\uu_0 & \uu_1 & \cdots & \uu_{s-1}
	\end{array}
	\right]
	G'_{\rm trunc}
	+
	\left[
	\begin{array}{cccc}
		\ee_0 & \ee_1 & \cdots & \ee_{s-1}
	\end{array}
	\right],
\end{equation}
where $G'_{\rm trunc}$ is given by the matrix
\begin{equation*}
	G'_{\rm trunc} =
	\scalebox{0.6}{$
		\left[
		\begin{array}{cccccccccccccccccccccccc}
			G_0' & G_1' & G_2' & G_3' & G_4' \\
			& G_0' & G_1' & G_2' & G_3' & G_4' \\
			&&\dd & \dd &\dd & \dd & \dd \\
			&&& \dd & \dd & \dd & \dd & \dd \\
			&&&& \dd & \dd & \dd & \dd & \dd \\
			&&&&& \dd & \dd & \dd & \dd & \dd \\
			&&&&&& G_0' & G_1' & G_2' & G_3' & G_4' \\
			&&&&&&& G_0' & G_1' & G_2' & G_3' & G_4' \\
			G_4' &&&&&&&& G_0' & G_1' & G_2' & G_3' \\
			G_3'& G_4'&&&&&&&& G_0' & G_1' & G_2' \\
			G_2'& G_3'& G_4' &&&&&&&& G_0' & G_1' \\
			G_1'& G_2'& G_3'& G_4' &&&&&&&& G_0'
		\end{array}
		\right]
		$}
	\in \mathbb{F}^{sk \times sn}.
\end{equation*}
Moreover, one can readily check that $G'_{\rm trunc}$ is the product of the three following matrices:
\[
S_{\rm trunc} =
\underbrace{
	\scalebox{0.6}{$
		\left[
		\begin{array}{cccccccccccccccccc}
			S_0 & S_1 & S_2 &&& \\
			& S_0 & S_1 & S_2 \\
			&& \dd & \dd & \dd \\
			&&& \dd & \dd & \dd \\
			&&&& S_0 & S_1 & S_2 \\
			S_2 &&&&& S_0 & S_1 \\
			S_1 & S_2 &&&&& S_0 \\
		\end{array}
		\right]
		$}
}_{\in \mathbb{F}^{sk \times sk}},
\hspace{0.15cm}
G_{\rm diag} =
\underbrace{
	\left[
	\begin{array}{cccc}
		G  \\
		& G \\
		&& \ddots \\
		&&& G
	\end{array}
	\right]
}_{\in \mathbb{F}^{sk \times sn}},
\text{ and }
\hspace{0.15cm}
P_{\rm trunc} =
\underbrace{
	\scalebox{0.6}{$
		\left[
		\begin{array}{ccccccccccccccccccccc}
			P_0 & P_1 & P_2 \\
			& P_0 & P_1 & P_2 \\
			&& \dd & \dd & \dd \\
			&&& \dd & \dd & \dd \\
			&&&& P_0 & P_1 & P_2 \\
			P_2 &&&&& P_0 & P_1 \\
			P_1 & P_2 &&&&& P_0 \\
		\end{array}
		\right]
		$}
}_{\in \mathbb{F}^{sn \times sn}}.
\]

Recall that the first step of the decryption consists in computing $\hat{\yy}(D) = \yy(D)T\DD \pmod{D^s-1}$, \ie, the vectors
\begin{equation}\label{eq:y_hat}
\hat{\yy}_{i} = \yy_{[i]_s} T_0 + \yy_{[i+1]_s} T_{-1} + \yy_{[i+2]_s} T_{-2}, \hspace{0.5cm} i \in \{ 0,1,\ldots,s-1 \}.
\end{equation}
So, if we define the matrix
\begin{equation*}
T_{\rm trunc} =
	\scalebox{0.6}{$
		\left[
		\begin{array}{ccccccccccccccccccccc}
			T_0 &&&&& T_{-2} & T_{-1} \\
			T_{-1} & T_0 &&&&& T_{-2} \\
			T_{-2} & T_{-1} & T_0 \\
			& T_{-2} & T_{-1} & T_0 \\
			&& \dd & \dd & \dd \\
			&&& T_{-2} & T_{-1} & T_0 \\
			&&&& T_{-2} & T_{-1} & T_0 \\
		\end{array}
		\right]
		$}
\in \mathbb{F}^{sn \times sn},
\end{equation*}
and denote by
\begin{align*}
\left[
\begin{array}{cccc}
    \hat{\uu}_0 & \hat{\uu}_1 & \cdots & \hat{\uu}_{s-1}
\end{array}
\right]
& =
\left[
\begin{array}{cccc}
    \uu_0 & \uu_1 & \cdots & \uu_{s-1}
\end{array}
\right]
S_{\rm trunc},
\\
\left[
\begin{array}{cccc}
    \hat{\ee}_0 & \hat{\ee}_1 & \cdots & \hat{\ee}_{s-1}
\end{array}
\right]
& =
\left[
\begin{array}{cccc}
    \ee_0 & \ee_1 & \cdots & \ee_{s-1}
\end{array}
\right]
T_{\rm trunc},
\end{align*}
then equalities \eqref{eq:y_hat} and \eqref{eq:ciphertext_trunc} imply that
\begin{align*}
	\left[
	\begin{array}{cccc}
		\hat{\yy}_0 & \hat{\yy}_1 & \cdots & \hat{\yy}_{s-1}
	\end{array}
	\right]
	& =
	\left[
	\begin{array}{cccc}
		\yy_0 & \yy_{1} & \cdots & \yy_{s-1}
	\end{array}
	\right] T_{\rm trunc} \\
	& =
	\left[
	\begin{array}{cccc}
		\uu_0 & \uu_1 & \cdots & \uu_{s-1}
	\end{array}
	\right]
	G'_{\rm trunc}T_{\rm trunc}
	+
	\left[
	\begin{array}{cccc}
		\ee_0 & \ee_1 & \cdots & \ee_{s-1}
	\end{array}
	\right]
	T_{\rm trunc} \\
	& =
	\left[
	\begin{array}{cccc}
		\hat{\uu}_0 & \hat{\uu}_1 & \cdots & \hat{\uu}_{s-1}
	\end{array}
	\right]
	G_{\rm diag}P_{\rm trunc}T_{\rm trunc}
	+
	\left[
	\begin{array}{cccc}
		\hat{\ee}_0 & \hat{\ee}_1 & \cdots & \hat{\ee}_{s-1}
	\end{array}
	\right].
\end{align*}

Now, the identity $P(D) T\DD = I_n$ gives us the relations
\begin{align*}
    P_0 T_{-2} & = 0, \\
    P_0 T_{-1} + P_1 T_{-2} & = 0, \\
    P_0 T_0 + P_1 T_{-1} + P_2 T_{-2} & = I_n, \\
    P_1 T_0 + P_2 T_{-1} & = 0, \\
    P_2 T_0 & = 0,
\end{align*}
so we have that
\[
P_{\rm trunc}
T_{\rm trunc}
=
\left[
    \begin{array}{cccc}
        I_n \\
        & I_n \\
        && \ddots \\
        &&& I_n
    \end{array}
\right] \in \mathbb{F}^{sn \times sn},
\]
and hence
\begin{align*}
\left[
\begin{array}{cccc}
    \hat{\yy}_0 & \hat{\yy}_1 & \cdots & \hat{\yy}_{s-1}
\end{array}
\right]
& =
	\left[
	\begin{array}{cccc}
		\hat{\uu}_0 & \hat{\uu}_1 & \cdots & \hat{\uu}_{s-1}
	\end{array}
	\right]
	G_{\rm diag}
	+
	\left[
	\begin{array}{cccc}
		\hat{\ee}_0 & \hat{\ee}_1 & \cdots & \hat{\ee}_{s-1}
	\end{array}
	\right] \\
& =
\left[
\begin{array}{cccc}
    \hat{\uu}_0 G + \hat{\ee}_0 & \hat{\uu}_1 G + \hat{\ee}_1 & \cdots & \hat{\uu}_{s-1} G + \hat{\ee}_{s-1}
\end{array}
\right].
\end{align*}
As each $\hat{\ee}_i$ is of the form $\hat{\ee}_{i} = \ee_{[i]_s} T_0 + \ee_{[i+1]_s} T_{-1} + \ee_{[i+2]_s} T_{-2}$, then Lemma \ref{lem:01} implies that $\wt(\hat{\ee}_i) \leq t$. Therefore, each block $\hat{\yy}_i = \hat{\uu}_i G + \hat{\ee}_i$ can be decoded by using an efficient decoding algorithm of the linear code generated by $G$ to compute each $\hat{\uu}_i$. Finally, we have that
\[
\left[
\begin{array}{cccc}
    \hat{\uu}_0 & \hat{\uu}_1 & \cdots & \hat{\uu}_{s-1}
\end{array}
\right]
=
\left[
\begin{array}{cccc}
    \uu_0 & \uu_1 & \cdots & \uu_{s-1}
\end{array}
\right] S_{\rm trunc},
\]
and since $S_{\rm trunc}$ is non singular, this is a determinate linear system of equations and the vectors $\uu_0, \uu_1, \ldots, \uu_{s-1}$ can be recovered.

\begin{remark}
It is easy to check that the scheme also works if the polynomial matrix $S(D)$ has larger degree. For the sake of reducing the size of the public key we stick to matrices $S(D)$ of degree 2.
\end{remark}

\subsection{Using Generalized Reed-Solomon codes}\label{conT}\label{subsection2.3}

The very general scheme described above must be correctly instantiated. In order to maximize the reduction of the size of the public key we shall propose the use of \emph{Generalized Reed-Solomon} (GRS) codes and build $S(D)$ and $T\DD$ in such a way that the resulting scheme is secure against structural attacks. In this section we provide a concrete class of such matrices $S(D)$ and $T\DD$.

\medskip

Recall that GRS codes are MDS codes and admit a generator matrix given by two vectors: $(\alpha_1,\ldots,\alpha_n) \in \FFF^n$, with $\alpha_i \neq \alpha_j$ for every $i \neq j$, and $(x_1,\ldots,x_n) \in (\FFF \setminus \{0\})^n$, defined as:
\begin{equation*}
G =
\left[
\begin{array}{cccc}
x_1 & x_2 & \cdots & x_n \\
x_1 \alpha_1 & x_2 \alpha_2 & \cdots & x_n \alpha_n \\
\vdots & \vdots & \ddots & \vdots \\
x_1 \alpha_1^{k-1} & x_2 \alpha_2^{k-1} & \cdots & x_n \alpha_n^{k-1}
\end{array}
\right].
\end{equation*}

\medskip

GRS codes are known for having a strong algebraic structure that it is difficult to hide. However, it was shown in \cite{GR2017,karan2018,karan2019} that weight two masks remove any identifiable algebraic structure of GRS codes, \ie, if $G$ is a generator matrix of a GRS code then $P$ can disguise the structure of $G$ in $GP$ if the columns of $P$ have at least weight two. For this reason, we impose that each nonzero column of $P_i$ has at least two nonzero elements (see details in Section \ref{sec:structural_attacks}).

\medskip

Further, it follows from Lemma \ref{lem:01} that the maximum weight of the rows of $T_i$, denoted by $\rho$, determines the number of errors one can add within an interval (see condition (\ref{eq:weight_e})). As we want to add as many errors as possible, we require the rows of the $T_i$'s to have the smallest possible weight, which in the construction proposed in this section is either zero or two. Hence, we impose the following conditions on $P(D)$ and $T\DD$:

\begin{Properties}\label{conditionsT} For each $i \in \{ 0,1,2 \}$,
\begin{description}
    \item[a)] each nonzero column of $P_i$ has at least two nonzero elements;
	\item[b)] each nonzero row of $T_{-i}$ has exactly two nonzero elements.
\end{description}
\end{Properties}

\medskip

Before presenting a large family of matrices $P(D)$ and $T\DD$ satisfying Properties \ref{conditionsT}, we recall a technical lemma about the determinant and inverse of a block matrix, first obtained by I. Schur \cite{Sc1917}.

\begin{lemma}{\cite[Formulas (2) and (4)]{Co1974}}\label{BlockM}
Let $T$ be a block matrix of the form
\[
T = \begin{sbmatrix}{cc}
A_{11} & A_{12} \\
A_{21} & A_{22}
\end{sbmatrix},\]
where $A_{11}$ and $A_{22}$ are non singular. Then,
\begin{description}
    \item[a)] $| T| = | A_{11} | \ | A_{22} - A_{21} A_{11}^{-1} A_{12}|$.
    \item[b)] If $T$ is non singular, the inverse of $T$ is
    \[T^{-1}=
    \begin{sbmatrix}{cc}
    (A_{11} - A_{12} A_{22}^{-1} A_{21})^{-1} & - A_{11}^{-1} A_{12} (A_{22} - A_{21} A_{11}^{-1} A_{12})^{-1}\\
    - A_{22}^{-1} A_{21} (A_{11} - A_{12} A_{22}^{-1} A_{21})^{-1} & (A_{22} - A_{21} A_{11}^{-1} A_{12})^{-1}
    \end{sbmatrix}.\]
\end{description}
\end{lemma}	

We propose a class of matrices $T\DD$ of the following form:

\begin{equation}\label{constructionT}
T\DD= \Pi\left[
	\begin{array}{c|r}
	A\DD & \beta A\DD  \\ \hline
	A\DD & A\DD
	\end{array}\right ]\in\FFF^{n \times n}\DDDD,
\end{equation}
with $n$ even, $\beta \notin \{0,1\}$, $\Pi \in \FFF^{n \times n}$ is a permutation matrix and the matrix $A=A\DD\in\FFF^{\frac{n}{2} \times \frac{n}{2}}\DDDD$ is randomly generated, satisfying the following conditions:

\begin{Properties}\label{propA}\
	\begin{enumerate}
		\item $A$ is an upper triangular matrix;
		\item The entries of the principal diagonal of $A$ are of the form $D^j$, with $j\in\{ -2,-1,0 \}$;
		\item Each row of $A$ has at most one entry of the form $\gamma D^j$ for each exponent $j\in\{-2,-1,0\}$, with $\gamma \in \mathbb{F} \setminus \{0\}$;
		\item All nonzero entries of a column of $A$ have the same exponent of $D$.
	\end{enumerate}	
\end{Properties}

%
%
%
%
%

\begin{theorem}\label{thm:0}
Let $|\mathbb{F}| > 2$. The matrices $T\DD$ described in (\ref{constructionT}) satisfy Properties \ref{conditionsT}.
\end{theorem}

\begin{proof} Since $\Pi$ only permutes the rows of
\[
	B\DD=\left [
	\begin{array}{c|r}
	A & \beta A  \\ \hline
	A & A
	\end{array}\right ]\in\FFF^{n \times n}\DDDD,
\]
then Properties \ref{conditionsT} {\bf b)} follows directly from Properties \ref{propA} {\bf 3)} and the block structure of $B\DD$. Furthermore, since $A$ is upper triangular, Properties \ref{propA} {\bf 2)} implies that $|A| = D^d$ for some integer $d$ and therefore $A^{-1}$ exists. Lemma \ref{BlockM} {\bf b)} implies that
\begin{equation}\label{eq:Tinv_block_structure}
T^{-1}\DD
=
B^{-1}\DD \Pi^{-1}
=
\begin{sbmatrix}{cc}
	\frac{1}{1-\beta}A^{-1} & \frac{\beta}{\beta-1}A^{-1}\\
	\frac{1}{\beta-1}A^{-1} & \frac{1}{1-\beta}A^{-1}
\end{sbmatrix} \Pi^{-1},
\end{equation}
and hence $T\DD$ is invertible. Notice that $\Pi^{-1}$ only permutes the columns of $B^{-1}\DD$, so it does not change the exponents of $D$.

\medskip

Properties \ref{propA} {\bf 4)} implies that $A$ can be decomposed as
\begin{equation*}
A
=
A'
\cdot
\left[
\begin{array}{cccc}
	D^{-i_1} &  \\
	& D^{-i_2} \\
	&& \ddots \\
	&&& D^{-i_{\frac{n}{2}}}
\end{array}
\right],
\end{equation*}
where $A'$ is the constant matrix obtained by taking the coefficients $\gamma$ of the monomials $\gamma D^i$ in the entries of $A$, and $i_1,\ldots,i_\frac{n}{2} \in \{ 0,1,2 \}$. Thus,
\[
A^{-1}
=
\left[
\begin{array}{cccc}
D^{i_1} &  \\
& D^{i_2} \\
&& \ddots \\
&&& D^{i_{\frac{n}{2}}}
\end{array}
\right] (A')^{-1},
\]
with $i_1,\ldots,i_{\frac{n}{2}} \in \{ 0,1,2 \}$ and hence $P(D) = T^{-1}\DD$ is of the form $P_0 + P_1 D + P_2 D^2$. Moreover, for each $j\in\{0,1,2\}$, the columns of $P_j$ will have an even number of nonzero entries because of the block structure in \eqref{eq:Tinv_block_structure}, so Properties \ref{conditionsT} {\bf a)} holds.
\end{proof}

\begin{example}\label{ex:TP}
In $\mathbb{F}_{13}$ the following matrix satisfies Properties \ref{propA}: 
	\[
	A =
    \scalebox{0.75}{$
	\begin{sbmatrix}{cccccc}
		D^{-1} & 9D^{-2} & 0 & 0 & 2 & 0 \\
		0 & D^{-2} & 2 & 0 & 0 & 0 \\
		0 & 0 & 1 & 9D^{-2} & 0 & 7D^{-1} \\
		0 & 0 & 0 & D^{-2} & 0 & 3D^{-1} \\
		0 & 0 & 0 & 0 & 1 & 6D^{-1} \\
		0 & 0 & 0 & 0 & 0 & D^{-1}
	\end{sbmatrix}
	$}
	=
    \scalebox{0.75}{$
	\begin{sbmatrix}{cccccc}
		1 & 9 & 0 & 0 & 2 & 0 \\
		0 & 1 & 2 & 0 & 0 & 0 \\
		0 & 0 & 1 & 9 & 0 & 7 \\
		0 & 0 & 0 & 1 & 0 & 3 \\
		0 & 0 & 0 & 0 & 1 & 6 \\
		0 & 0 & 0 & 0 & 0 & 1
	\end{sbmatrix}
	$}
    \scalebox{0.75}{$
	\begin{sbmatrix}{cccccc}
		D^{-1} & 0 & 0 & 0 & 0 & 0 \\
		0 & D^{-2} & 0 & 0 & 0 & 0 \\
		0 & 0 & 1 & 0 & 0 & 0 \\
		0 & 0 & 0 & D^{-2} & 0 & 0 \\
		0 & 0 & 0 & 0 & 1 & 0 \\
		0 & 0 & 0 & 0 & 0 & D^{-1}
	\end{sbmatrix}	
	$}
	.
	\]
Take  $\Pi = I_{12}$ the identity matrix of size $12$, and $\beta=5$. Then, by Theorem \ref{thm:0}, we have that
	\[
	T\DD =\left [
	\begin{array}{c|c}
	A & \beta A  \\ \hline
	A & A
	\end{array}\right ]=
    \scalebox{0.75}{$
	\begin{sbmatrix}{cccccc|cccccc}
		D^{-1} & 9D^{-2} & 0 & 0 & 2 & 0 & 5D^{-1} & 6D^{-2} & 0 & 0 & 10 & 0 \\
		0 & D^{-2} & 2 & 0 & 0 & 0 & 0 & 5D^{-2} & 10 & 0 & 0 & 0 \\
		0 & 0 & 1 & 9D^{-2} & 0 & 7D^{-1} & 0 & 0 & 5 & 6D^{-2} & 0 & 9D^{-1} \\
		0 & 0 & 0 & D^{-2} & 0 & 3D^{-1} & 0 & 0 & 0 & 5D^{-2} & 0 & 2D^{-1} \\
		0 & 0 & 0 & 0 & 1 & 6D^{-1} & 0 & 0 & 0 & 0 & 5 & 4D^{-1} \\
		0 & 0 & 0 & 0 & 0 & D^{-1} & 0 & 0 & 0 & 0 & 0 & 5D^{-1} \\ \hline
		D^{-1} & 9D^{-2} & 0 & 0 & 2 & 0 & D^{-1} & 9D^{-2} & 0 & 0 & 2 & 0 \\
		0 & D^{-2} & 2 & 0 & 0 & 0 & 0 & D^{-2} & 2 & 0 & 0 & 0 \\
		0 & 0 & 1 & 9D^{-2} & 0 & 7D^{-1} & 0 & 0 & 1 & 9D^{-2} & 0 & 7D^{-1} \\
		0 & 0 & 0 & D^{-2} & 0 & 3D^{-1} & 0 & 0 & 0 & D^{-2} & 0 & 3D^{-1} \\
		0 & 0 & 0 & 0 & 1 & 6D^{-1} & 0 & 0 & 0 & 0 & 1 & 6D^{-1} \\
		0 & 0 & 0 & 0 & 0 & D^{-1} & 0 & 0 & 0 & 0 & 0 & D^{-1}	
	\end{sbmatrix}
    $}
	\]
	and
\[
	P(D)=T^{-1}\DD =
    \scalebox{0.75}{$
    \begin{sbmatrix}{cccccc|cccccc}
		3 D & 12 D & 2 D & 8 D & 7 D & 11 D & 11 D & 5 D & 3 D & 12 D & 4 D & 10 D \\
		0 & 3 D^2 & 7 D^2 & 2 D^2 & 0 & 10 D^2 & 0 & 11 D^2 & 4 D^2 & 3 D^2 & 0 & 2 D^2 \\
		0 & 0 & 3 & 12 & 0 & 8 & 0 & 0 & 11 & 5 & 0 & 12 \\
		0 & 0 & 0 & 3 D^2 & 0 & 4 D^2 & 0 & 0 & 0 & 11 D^2 & 0 & 6 D^2 \\
		0 & 0 & 0 & 0 & 3 & 8 & 0 & 0 & 0 & 0 & 11 & 12 \\
		0 & 0 & 0 & 0 & 0 & 3 D & 0 & 0 & 0 & 0 & 0 & 11 D \\ \hline
		10 D & D & 11 D & 5 D & 6 D & 2 D & 3 D & 12 D & 2 D & 8 D & 7 D & 11 D \\
		0 & 10 D^2 & 6 D^2 & 11 D^2 & 0 & 3 D^2 & 0 & 3 D^2 & 7 D^2 & 2 D^2 & 0 & 10 D^2 \\
		0 & 0 & 10 & 1 & 0 & 5 & 0 & 0 & 3 & 12 & 0 & 8 \\
		0 & 0 & 0 & 10 D^2 & 0 & 9 D^2 & 0 & 0 & 0 & 3 D^2 & 0 & 4 D^2 \\
		0 & 0 & 0 & 0 & 10 & 5 & 0 & 0 & 0 & 0 & 3 & 8 \\
		0 & 0 & 0 & 0 & 0 & 10 D & 0 & 0 & 0 & 0 & 0 & 3 D
	\end{sbmatrix}
    $}
\]
satisfy Properties \ref{conditionsT}.
\end{example}

%

\medskip


For the construction of $S(D) = S_0 + S_1 D + S_2 D^2$ we only require $S_{\rm trunc}$ to be invertible. This loose condition on the coefficients of $S(D)$ allows to add a lot of randomness into the system.

\section{Attacks Against the Proposed Cryptosystem}\label{sec:security}

In this section, we investigate possible attacks to the proposed cryptosystem and show why these attacks are expected to fail. The results of this section show that the proposed cryptosystem provides increased key security with respect to previous variants of the McEliece cryptosystem, in such a way as to allow the use of GRS codes without incurring in the attacks that have prevented their use up to now. We divide the section into two parts according to the two main classes of attacks to the McEliece-type cryptosystems: plaintext recovery attacks and structural attacks.

\subsection{Plaintext recovery}\label{Section ISD}

This is the most general attack procedure against code-based cryptosystems and aims to decode the ciphertext without requiring any knowledge of the private key. Trying to decode directly a convolutional code using maximum-likelihood decoding or sequential decoding seems very difficult due to the large number of Trellis transitions.

\medskip

Plaintext recovery attacks are typically performed using information set decoding algorithms (ISD), first introduced by Prange in \cite{prange}.
The ISD algorithms aim to solve the following NP-hard problem: Given the generator matrix $G'\in\FFF^{K \times N}$ of an arbitrary linear block code and a vector $\yy= \textbf{u}G' + \textbf{e}$, recover $\uu$. The first step of any ISD algorithm is to find a size-$K$ index set $I\subset \{1,2,\dots,N\}$ such that the submatrix of $G'$ with the columns indexed by $I$ is a non singular matrix of order $K$. The set $I$ is called an {\em Information Set}. The second step depends on the algorithm we are using, but the basic idea is to guess the $I$-indexed part $\ee_I$ of the error vector $\ee$ according to a predefined method (that depends on each specific algorithm) and try to obtain the whole $\ee$ from these assumptions.



\medskip

An example of these ISD algorithms is the Stern's ISD algorithm. A detailed description for codes over arbitrary finite fields can be found in \cite[Sec. 3]{Peters2010} by C. Peters, where she also presents a work factor (WF) estimation, \ie, the estimated number of operations the algorithm performs until the message is recovered: for $\FFF$ with $q$ elements, where $q$ is prime, $T$ the number of errors and the parameters $p$ and $\ell$, the work factor of this ISD algorithm is given by
\begin{equation}\label{wffinal}
\WF_{p,\ell}(q,N,K,T)
=
\mathcal{S}_{p,\ell}
\frac{\binom{N}{T}}{\binom{N-K-\ell}{T-2p}\binom{\frac{K}{2}}{p}^2},
\end{equation}
where
\begin{align*}
\mathcal{S}_{p,\ell} & = (N-K)^2(N+K)+\ell \left(\frac{K}{2}-p+1+ 2 \binom{\frac{K}{2}}{p} (q-1)^p \right)\\
& \hspace{0.5cm} + \frac{2pq(T-2p+1)(2q-3)(q-1)^{2p-2}}{q^\ell}\binom{\frac{K}{2}}{p}^2.
\end{align*}

Parameters $p$ and $\ell$ have to be determined to minimize $\WF_{p,\ell}(q,N,K,T)$. A few improvements were made to this algorithm (see for example \cite{FS2009} and \cite{Irterlando2020}) but as it was pointed out in \cite{GR2017} these new ISD algorithms will not significantly decrease the work factor.

\medskip

The attacks based on ISD algorithms will be analyzed considering three different situations: when the attacker deals with the whole message without exploiting the block structure of $G'_{\rm trunc}$, when only intervals of the sequence that forms the ciphertext are considered and when the block structure of $G'_{\rm trunc}$ is exploited.

\medskip

\subsubsection{ISD attacks on the full matrix $G'_{\rm trunc}$}\label{sec:attack_full_matrix}

An attacker could consider
\begin{equation*}
\left[
\begin{array}{cccc}
    \yy_0 & \yy_1 & \cdots & \yy_{s-1}
\end{array}
\right]
=
\left[
\begin{array}{cccc}
    \uu_0 & \uu_1 & \cdots & \uu_{s-1}
\end{array}
\right]
G'_{\rm trunc}
+
\left[
\begin{array}{cccc}
    \ee_0 & \ee_1 & \cdots & \ee_{s-1}
\end{array}
\right]
\end{equation*}
and perform an attack using the Stern's ISD algorithm. In this case the work factor is given by \eqref{wffinal} with $K=sk$, $N=sn$ and $T \approx \frac{st}{6}$, which depend on $s$, so this attack can be countered by taking larger values for $s$.

\medskip

\medskip

\subsubsection{Sequential plaintext recovery attacks}\label{sec:sequential attacks}

The main difference between block and convolutional codes, is that convolutional encoders may have different states and after encoding an input we possibly move into another state. Hence, the information vector encoded at each instant depends on the previous information vectors, so it seems natural to try to attack the first data received. However, since $\yy(D) = \uu(D) G'(D) + \ee(D)$ is computed modulo the polynomial $D^s-1$ we have that
\[
\yy_i =
\uu_{[i]_s} G'_0 + \uu_{[i-1]_s} G'_1 + \cdots + \uu_{[i-4]_s} G'_4 + \ee_i,
\]
for all $i \in \{0,1,\ldots,s-1\}$, so each $\yy_i$ depends on five vectors $\uu_i$, which makes difficult to compute any $\ee_i$, even at the first instants.

\medskip

We consider that the cryptanalysis has been successful if the attacker is able to recover a part of the message. Hence, the attacker can consider an interval, for example $\left[ \yy_0 \ \yy_{1} \ \cdots \ \yy_{r} \right]$ with $r \in \{0,1,\ldots,s-1\}$, and attack it with a standard ISD attack. In this case, we obtain a linear system
a system of $(r+1)n$ equations and $\min \{ (r+5)k, sk \}$ unknowns. In order to simplify the analysis we assume that Alice adds $\frac{t}{6} = \frac{n-k}{12}$ errors at each instant instead of $\frac{t}{2} = \frac{n-k}{4}$ in each interval $[\ee_{[i]_s} \ \ee_{[i+1]_s} \ \ee_{[i+2]_s}]$. This approach is conservative in the sense that we assume that such particular distribution of errors always occur, which is very seldom the case. Thus, the expected cost of this step when using the Stern's ISD given by \eqref{wffinal} is
\begin{equation}\label{eq:wf_interval}
	\WF_{p,\ell} \left( q, (r+1)n, \min \{ (r+5)k, sk \}, \frac{(r+1)(n-k)}{12}\right).
\end{equation}
Hence the attacker must select $r$, $p$ and $\ell$ in order to minimize this value. Notice that the attack in Section \ref{sec:attack_full_matrix} is the attack explained above when $r=s-1$.

\medskip

\subsubsection{ISD using the block structure}\label{sec:optimized_ISD}

A necessary condition for $I$ to be an information set is that the submatrix formed by the columns indexed by $I$ is non singular. In the above attack it is assumed implicitly that this is always the case. However, due to the structure of the blocks of zeros in $G'_{\rm trunc}$ this obviously cannot happen for some selection of columns.

\medskip

Here we explain how one can use the block structure of $G'_{\rm trunc}$ to get an attack. For simplicity, we show how to adapt the Prange ISD attack. Other ISD attacks can be adapted as well in a similar way. In the Prange ISD attack we consider that the information set $I$ has been chosen successfully if it does not contain any index where an error has been added, that is, if $I \cap \supp([ \ee_0 \ \ee_1 \ \cdots \ \ee_{s-1} ]) = \emptyset$. Again, in order to simplify the analysis we assume that Alice adds $\frac{t}{6} = \frac{n-k}{12}$ errors at each instant.





\medskip

In the attack we propose here the attacker selects the same number of columns in each column block to form $I$, \ie, $I = \bigcup_{i = 0}^{s-1} I_i $, where $I_j$ is associated to the column block $j$, and all $I_j$ have size $k$. Since the number of errors per block is assumed to be $\frac{t}{6}$, the probability of success for a random choice of indices $I_j$ is less than or equal to
\[
\frac{\binom{n-\frac{t}{6}}{ k }}{\binom{n}{ k }}.
\]

%
%

\noindent and therefore, the probability $\mathcal{P}$ of selecting an information set $I$ for $G'_{\rm trunc}$ whose indices are not associated to error positions and such that the columns indexed by $I$ form an invertible matrix is
\begin{equation}\label{eq:probability}
\mathcal{P}
\leq
\frac{\binom{n-\frac{t}{6}}{k}^s}{\binom{n}{ k }^s} .
\end{equation}
Once the set $I$ has been chosen, the attacker solves the linear system $\yy_I = \vect{x} (G'_{\rm trunc})_I$, where $\yy_I$ is formed by the components of $[ \yy_0 \ \yy_1 \ \cdots \ \yy_{s-1} ]$ indexed by $I$ and $(G'_{\rm trunc})_I$ is the matrix obtained by taking the columns of $G'_{\rm trunc}$ indexed by $I$. The cost of solving this linear system is about $(sk)^3$. Therefore the expected work factor can be estimated with the formula
\begin{equation}\label{wffinal1}
\WF = (sk)^3 \cdot \mathcal{P}^{-1}.
\end{equation}


For simplicity, we consider that the inequality \eqref{eq:probability} holds with equality when estimating the $\WF$ of this attack. Table \ref{tabex} contains the work factors \eqref{wffinal}, \eqref{eq:wf_interval} and \eqref{wffinal1} for certain parameters. In all the cases, the work factor obtained using \eqref{eq:wf_interval} is lower or equal (when $r=s-1$) than the work factors obtained using \eqref{wffinal} or \eqref{wffinal1}, so it will be the one taken into account when proposing the parameters for $128$-bit, $256$-bit and $512$-bit security level against these attacks despite this $\WF$ comes from the strong assumption that every $(r+5)k$ columns (or $sk$ columns, depending on $r$) of $G'_{\rm trunc}$ form an invertible matrix.

\medskip

Finally, it is important to note that the size of $(G'_{\rm trunc})_I$ depends on $s$, that can be arbitrarily large without increasing the size of the key.

\subsection{Structural attacks}\label{sec:structural_attacks}

Structural attacks are more specific attacks that aim at exploiting the particular structure of the keys. Obviously, the scheme is broken if the public key $G'(D)$ can be factorized into the secret matrices $S(D)$, $G$ and $P(D)$. In our scheme this triplet is not unique in the sense that it is enough to find any matrices $\mathcal{S}(D)$, $\mathcal{G}$ and $\mathcal{P}(D)$ such that $G'(D) = \mathcal{S}(D) \mathcal{G} \mathcal{P}(D)$ and such that the decoding method explain in Section \ref{subsection2.1} can be performed.

\medskip

If one consider the code generated by $\mathcal{G}=U G \Delta \Gamma$, with $U \in \mathbb{F}^{k \times k}$ non singular, $\Delta \in \mathbb{F}^{n \times n}$ non singular diagonal  and $\Gamma \in \mathbb{F}^{n \times n}$ a permutation matrix, then, the triplet
\[ \{ \mathcal{S}(D)=S(D) U^{-1},\ \mathcal{G}=U G \Delta \Gamma,\ \mathcal{P}(D)=( \Delta \Gamma)^{-1} P(D) \} \]
can be used to decode the ciphertext.

\medskip

\subsubsection{Brute force attacks}

Naive brute force attacks aiming to find $ \mathcal{S}(D),\ \mathcal{G},\ \mathcal{P}(D)$ seem, \emph{a priori}, unfeasible due to the large amount of matrices an attacker has to check.

\medskip

An attacker looks for a matrix $\mathcal{S}(D)$ such that there exists $U \in \mathbb{F}^{k \times k}$ non singular and $\mathcal{S}(D) U = S(D)$. We define the following equivalence relation
\begin{align*}
    M(D) \sim N(D) & \text{ if and only if there exists $U \in \mathbb{F}^{k \times k}$ non singular such that } M(D) = N(D) U.
\end{align*}
Let us denote by $[M(D)]$ the equivalence class of a matrix $M(D)$. Thus, any matrix $\mathcal{S}(D) \in [S(D)]$ is a valid candidate for the attacker, so the attacker can check only one element per class $[M(D)]$. Lemma \ref{lem:count_S} gives us a lower bound on number of classes $[M(D)]$. Before, we need to prove an auxiliary result.




\begin{lemma}\label{lem:count_S_1} Assume than $s<q$. Then the number of matrices $M(D) = I_k + M_1 D + M_2 D^2$ with a non singular $M_{\rm trunc} \in \FFF^{sk \times sk}$ is lower bounded by
\begin{equation}\label{eq:bound_S}
(q-s)^k
\prod_{j=1}^k \left( q^k - q^{j-1} \right).
\end{equation}
\end{lemma}

\begin{proof} The number of solutions of the equation $x^s-1=0$ in $\FFF$ is less than or equal to $s$, so the number of matrices
\[
\Delta =
\left[
    \begin{array}{cccc}
         \lambda_1 \\
        & \lambda_2 \\
        && \ddots \\
        &&& \lambda_k
    \end{array}
\right],
\]
where $\lambda_1,\lambda_2,\ldots,\lambda_k$ are not solutions of $x^s-1=0$ is greater than or equal to $(q-s)^k$. For each of those matrices we can consider any non singular matrix $Q \in \FFF^{k \times k}$. The number of non singular matrices in $\FFF^{k \times k}$ is given by $\prod_{j=1}^k \left( q^k - q^{j-1} \right)$. With these two matrices we define $M^* = -Q \Delta Q^{-1}$. Then the matrix $M(D) = I_k + M^* D + 0 D^2$ gives us an $M_{\rm trunc} \in \FFF^{sk \times sk}$ of the form
\[
M_{\rm trunc}
=
\left[
\begin{array}{cccccccccccccccccc}
     I_k & M^* \\
    & I_k & M^* \\
    && \ddots & \ddots \\
    &&& \ddots & M^* \\
    &&&& I_k & M^* \\
    M^* &&&&& I_k \\
\end{array}
\right] \in \FFF^{sk \times sk}.
\]
If we subtract to the last row block the first row block multiplied by $M^*$, then the second row block multiplied by $-(M^*)^2$, then the third row block multiplied by $(M^*)^3$, and so on, we can transform the above matrix into
\[
\left[
\begin{array}{cccccccccccccccccc}
     I_k & M^* \\
    & I_k & M^* \\
    && \ddots & \ddots \\
    &&& \ddots & M^* \\
    &&&& I_k & M^* \\
    &&&&& I_k - (-1)^s(M^*)^s
\end{array}
\right].
\]
The determinant of this matrix is $|I_k-(-1)^s (M^*)^s| = |I_k - Q \Delta^s Q^{-1}|$ so it is different from zero if and only if all the eigenvalues of $Q \Delta^s Q^{-1}$ are different from 1. This is equivalent to say that all the eigenvalues of $\Delta^s$ are different from 1, and this happens if and only if none of the elements $\lambda_1,\lambda_2,\ldots,\lambda_k$ satisfy the equation $x^s=1$, which is true by construction.
Therefore, $M_{\rm trunc}$ is non singular. Hence, the result follows from the fact that the number of matrices of the particular type $M(D) = I_k + M^* D + 0 D^2$ is lower bounded by
\[
(q-s)^k
\prod_{j=1}^k \left( q^k - q^{j-1} \right).
\]

\end{proof}

\begin{lemma}\label{lem:count_S} Assume that $s<q$. Then the number of classes $[M(D)]$ where $M(D) = M_0 + M_1 D + M_2 D^2$ with a non singular $M_{\rm trunc}\in \FFF^{sk \times sk}$ is lower bounded by \eqref{eq:bound_S}.
\end{lemma}

\begin{proof} The set of matrices $M(D) = M_0 + M_1 D + M_2 D^2$ with both $M_0$ and $M_{\rm trunc}\in \FFF^{sk \times sk}$ non singular forms a subset of the matrices $M(D) = M_0 + M_1 D + M_2 D^2$ with a non singular $M_{\rm trunc}\in \FFF^{sk \times sk}$. Since for the former set each class $[M(D)]$ admits a unique representative of the form $M'(D) = I_k + M_1' D + M_2' D^2$, the result is deduced from Lemma \ref{lem:count_S_1}.
\end{proof}



Table \ref{tab:number_of_Ss} illustrates the magnitude of the naive lower bound on the number of classes $[M(D)]$ given by Lemma \ref{lem:count_S} for the parameters that will be proposed in Table \ref{tabex}.

\begin{table}[h]
	\begin{center}
		\renewcommand{\arraystretch}{1.2}
		\begin{tabular}{|c|c|c|c|c|}\hline
			$q$ & $k$ & $s$ & $(q-s)^k \prod_{j=1}^k ( q^k - q^{j-1})$ \\ \hline
			127 & 66 & 30 & $2^{30878.29}$ \\
			127 & 72 & 29 & $2^{36705.59}$ \\
			127 & 72 & 24 & $2^{36710.76}$ \\
			251 & 142 & 28 & $2^{161845.93}$ \\
			251 & 148 & 25 & $2^{175766.07}$ \\
			251 & 160 & 22 & $2^{205325.78}$ \\
			509 & 288 & 29 & $2^{748357.97}$ \\
			509 & 288 & 28 & $2^{748358.83}$ \\
			509 & 300 & 28 & $2^{811909.93}$ \\
			\hline	
    		\end{tabular}
	\end{center}
	\caption{Lower bound on the number of classes $[M(D)]$}
	\label{tab:number_of_Ss}
\end{table}

\medskip

Now, we focus on the secret matrix $P(D)$. An attacker searches for any matrix $\mathcal{P}(D)$ such that there exist non singular matrices $\Delta, \Gamma \in \FFF^{n \times n}$ with $\Delta$ being a diagonal matrix, $\Gamma$ a permutation matrix and $\Delta \Gamma \mathcal{P}(D) = P(D)$. Notice that this would imply that $\mathcal{P}^{-1}(D) = T\DD \Delta \Gamma$, and hence, by (\ref{constructionT}),
\[
T\DD
=
\Pi
\left [
    \begin{array}{c|r}
	A & \beta A \\ \hline
	A & A
    \end{array}
\right ] \Delta \Gamma
=
\Pi
\left [
    \begin{array}{c|r}
	A' & \beta A' \\ \hline
	A' & A'
    \end{array}
\right ] \Gamma,
\]
for some upper triangular matrix $A'$. Recall that the entries in the main diagonal of $A$ are of the form $D^j$ and this is known by the attacker, so if $\Delta \neq I_n$ then the entries in the main diagonal of the blocks $A'$ would be of the form $aD^j$ with some $a$'s different from 1 and it does not seem a great advantage to consider this case. On the other hand, $\Gamma$ permutes the columns, but the attacker already knows the upper triangular structure of $A$, so it does not seem a great advantage to consider the case $\Gamma \neq I_n$, so we consider the situation when the attacker tries to find $\mathcal{P}(D) = P(D)$. We restrict our attention to find $A$ and $\beta$.

\medskip

\begin{notation} For each $i \in \{-2,-1,0\}$ the number of columns of $A\DD$ having $D^i$ is denoted as $d_i$.
\end{notation}

\begin{lemma}\label{lem:num_T} Let $A \in \FFF^{\frac{n}{2} \times \frac{n}{2}}[D^{-1},D]$ fulfilling Properties \ref{propA}. Let $i^* \in \{-2,-1,0\}$ be the exponent of $D$ in the entry $a_{\frac{n}{2},\frac{n}{2}}$, and let $i^{**} \in \{-2,-1,0\} \setminus \{i^{*}\}$ be the exponent of $D$ of the last element in the main diagonal which is not $D^{i^*}$. Then the number of matrices $T\DD$ defined above is at least $(q-2) \cdot q^{n-2d_{i^*}-d_{i^{**}}}$.
\end{lemma}

\begin{proof}The last column of $A$ only contains elements of the form $\gamma D^{i^*}$ with $\gamma \in \FFF$. Since each row of $A$ cannot have more than one entry with $D^{i^*}$, and $d_{i^*}$ of them are in the main diagonal, there are $\frac{n}{2}-d_{i^*}$ entries in that column (excluding $a_{\frac{n}{2},\frac{n}{2}}$) that can have an element of the form $\gamma D^{i^*}$. Thus, for the last column we have $q^{\frac{n}{2}-d_{i^*}}$ possibilities.

\medskip

Let $(j^{**},j^{**})$ be the indices of the last entry of the main diagonal which is not $D^{i^*}$. Then $j^{**}\geq \frac{n}{2}-d_{i^*}$. This means, that the corresponding column has at least $\left (\frac{n}{2}-1\right )-d_{i^*}$ entries above the main diagonal. Since each row of $A$ cannot have more than one entry with $D^{i^{**}}$, and $d_{i^{**}}$ of them are in the main diagonal, then there are at least $\left (\frac{n}{2}-1\right )-d_{i^*}-(d_{i^{**}}-1)$ entries in that columns above the main diagonal that can have an element of the form $\gamma D^{i^{**}}$. Thus, for that column we have at least $q^{\frac{n}{2}-d_{i^*}-d_{i^{**}}}$ possibilities.

\medskip

Since $\beta \in \FFF \setminus \{ 0,1 \}$, we have at least $(q-2) \cdot q^{n-2d_{i^*}-d_{i^{**}}}$ possibilities for $T\DD$.
\end{proof}

Table \ref{tab:number_of_Ts} shows the magnitude of the lower bound on the number of matrices $T\DD$ given by Lemma \ref{lem:num_T} for certain values of $(d_{-2},d_{-1},d_0)$.

\begin{table}[h]
	\begin{center}
		\renewcommand{\arraystretch}{1.2}
		\begin{tabular}{|c|c|c|c|}\hline
			$q$ & $n$ & $(d_{-2},d_{-1},d_0)$ & $(q-2) \cdot q^{n-2d_{i^*}-d_{i^{**}}}$ \\ \hline
			127 & 90 & $(15,15,15)$ & $2^{321.46}$ \\
			127 & 96 & $(16,16,16)$ & $2^{342.42}$ \\
			127 & 108 & $(18,18,18)$ & $2^{384.35}$ \\
			251 & 202 & $(34,33,34)$ & $2^{805.11}$ \\
			251 & 220 & $(37,36,37)$ & $2^{876.86}$ \\
			251 & 244 & $(41,40,41)$ & $2^{972.52}$ \\
			509 & 396 & $(66,66,66)$ & $2^{1789.31}$ \\
			509 & 408 & $(68,68,68)$ & $2^{1843.26}$ \\
			509 & 420 & $(70,70,70)$ & $2^{1897.21}$ \\ \hline
		\end{tabular}
	\end{center}
	\caption{Lower bound on the number of matrices $\mathcal{P}(D)$}
	\label{tab:number_of_Ts}
\end{table}


Hence, naive brute force approaches to recover a valid triplet $\{ \mathcal{S}(D), \mathcal{G}, \mathcal{P}(D) \}$ seem unfeasible. Therefore better structural attacks should be considered.

\medskip

\subsubsection{Attacks based on the Smith normal form}

An interesting idea could be to compute the Smith normal form of $G'(D)$ (see \cite[Sec. 2.2]{JohannessonZigangirov2015}, ). This gives us a factorization
\[ G'(D) = X(D) \Gamma(D) Y(D), \]
where $\Gamma(D)$ is known as the \emph{Smith normal form} of $G'(D)$ and fulfills certain properties. It happens that if $\Gamma(D) = [I_k \mid 0]$, then the attacker can choose the generator matrix $C$ of some code having an efficient decoding algorithm (not necessarily a GRS code) and try to find a factorization $C = M [I_k \mid 0] N$, with $M \in \mathbb{F}^{k \times k}$ and $N \in \mathbb{F}^{n \times n}$ invertible, such that $Y^{-1}(D)N$ is a polynomial matrix with each coefficient matrix having few nonzero elements in each row. If that is the case, then the attacker has found a factorization of $G'(D)$:
\[ G'(D) = \underbrace{X(D)M^{-1}}_{\mathcal{S}(D)} \underbrace{C}_{\mathcal{G}} \underbrace{N^{-1}Y(D)}_{\mathcal{P}(D)}, \]
that can be used to decrypt using a similar procedure as the one explained in Section \ref{sec:decryption}. However, for this attack to work the attacker needs to obtain $\Gamma(D) = [I_k \mid 0]$ and at most weight two in the rows of the inverse of $\mathcal{P}(D)$ if $C$ is the generator matrix of a GRS code. After computing numerous examples these conditions were far from being satisfied even for very small parameters. See Example \ref{ex:smith_form} for an instance of such decomposition for small parameters.  

\begin{example}\label{ex:smith_form}
	Over $\FFF_7$ we consider the following matrices to be part of the private key
	\begin{align*}
		S(D) & =
		\left[
		\scalebox{0.75}{$
			\begin{array}{rrrr}
				2 + D^2 & 2 + 5D & 5 & 4 + 4D + 2D^2 \\
				3 + 5D + 5D^{2} & 3 + 4D & 1 + 6D & 6 + 4D + 4D^2 \\
				4 + D + D^2 & 5 + 4D & 2 + 5D + 4D^2 & 5 + 4D + 2D^2 \\
				1 + D + D^2 & 6 + 4D + 4D^2 & 1 + 5D^2 & 3D + 4D^2
			\end{array}
			$}
		\right],
		&
		G & =
		\left[
		\scalebox{0.75}{$
			\begin{array}{rrrrrr}
				4 & 6 & 1 & 1 & 2 & 3 \\
				6 & 5 & 1 & 0 & 6 & 4 \\
				2 & 3 & 1 & 0 & 4 & 3 \\
				3 & 6 & 1 & 0 & 5 & 4
			\end{array}
			$}
		\right],
	\end{align*}
	\begin{align*}
		T\DD =
		\left[
		\scalebox{0.75}{$
		\begin{array}{rrrrrr}
			D^{-2} & 3D^{-1} & 2 & D^{-2} & 3D^{-1} & 2 \\
			D^{-2} & 3D^{-1} & 2 & 3D^{-2} & 2D^{-1} & 6 \\
			0 & 0 & 1 & 0 & 0 & 3 \\
			0 & D^{-1} & 0 & 0 & 3D^{-1} & 0 \\
			0 & D^{-1} & 0 & 0 & D^{-1} & 0 \\
			0 & 0 & 1 & 0 & 0 & 1
		\end{array}
		$}
		\right].
	\end{align*}
For all $3 \leq s \leq 100$ the matrix $S_{\rm trunc}$ is non singular except when $s \in \{ 16,32,48,64,80,96 \}$. We have that
	\begin{align*}
		P(D) & =
		\left[
		\scalebox{0.75}{$
			\begin{array}{rrrrrr}
				5 D^2 & 3 D^2 & D^2 & 5 D^2 & 6 D^2 & 4 D^2 \\
				0 & 0 & 0 & 3 D & 5D & 0 \\
				0 & 0 & 3 & 0 & 0 & 5 \\
				3 D^2 & 4 D^2 & 6 D^2 & 2 D^2 & 5 D^2 & D^2 \\
				0 & 0 & 0 & 4 D & 3 D & 0 \\
				0 & 0 & 4 & 0 & 0 & 3
			\end{array}
			$}
		\right],
	\end{align*}
so the public key is
	\begin{align*}
		& G'(D) = \\
		& \left[
		\scalebox{0.75}{$
			\scriptstyle
			\begin{array}{rrrrrr}
				6D^{2} + 4D^{4} & D^{2} + 6D^{4} & 2 + 3D + 2D^{2} + 2D^{4} & D + D^{2} + 4D^{3} + 3D^{4} & 4D + 6D^{2} + 2D^{4} & 4 + 6D + D^{2} + 6D^{4} \\
				5D^{3} & D^{2} + 4D^{4} & 2D + 2D^{2} + 6D^{4} & 4D^{2} + 2D^{3} + 2D^{4} & 2D + D^{2} + 2D^{3} & 6 + 3D + 5D^{2} + 4D^{3} \\
				D^{2} + D^{3} + 2D^{4} & D^{2} + 2D^{4} & 4D + 6D^{2} + 3D^{4} & D^{2} + 6D^{3} + D^{4} & D + 2D^{2} + D^{4} & 2 + 3D + 4D^{2} + 5D^{3} + 3D^{4} \\
				3D^{2} + 6D^{3} + D^{4} & 4D^{2} + 3D^{3} & 4 + D + 3D^{2} + D^{3} & 5D + 4D^{2} + 2D^{3} & 6D + 5D^{2} + D^{3} + 4D^{4} & 4 + 4D^{2} + 2D^{3} + 5D^{4}
			\end{array}
			$}
		\right].
	\end{align*}
	The Smith normal form of this matrix is
	\begin{align*}
		\Gamma(D) =
		\left[
		\scalebox{0.75}{$
			\begin{array}{cccccc}
				1 & 0 & 0 & 0 & 0 & 0 \\
				0 & 1 & 0 & 0 & 0 & 0 \\
				0 & 0 & D & 0 & 0 & 0 \\
				0 & 0 & 0 & D^{9} + 2 D^{8} + 5 D^{6} + 2 D^{3} + 4 D^{2} + 3 D & 0 & 0
			\end{array}
			$}
		\right]
	\end{align*}
	with
	\begin{align*}
		& X(D) = \\
		& \left[
		\scalebox{0.7}{$
			\begin{array}{rrrr}
				D^{5} + 6 D^{4} + 5 D^{3} + D^{2} + 5 D + 1 & 5 & 0 & 0 \\
				6 & 0 & 0 & 0 \\
				4 D^{5} + 4 D^{4} + 2 D^{3} + 2 D^{2} + 4 D + 2 & 2 D^{7} + 5 D^{6} + 3 D^{5} + 3 D^{4} + 5 D^{3} + 3 D^{2} + 4 D & 2 D^{8} + 6 D^{7} + D^{6} + 2 D^{5} + 3 D^{3} + 4 D^{2} + 4 D & 4 \\
				2 D^{5} + 5 D^{4} + 2 D^{3} + D^{2} + 2 D + 5 & 4 D^{7} + 5 D^{6} + 4 D^{5} + D^{4} + 3 D^{3} + 2 D^{2} + 4 D + 3 & 4 D^{8} + 5 D^{7} + 2 D^{6} + 4 D^{5} + 6 D^{3} + D^{2} + D + 3 & 1
			\end{array}
			$}
		\right]
	\end{align*}
	and $Y(D) = [ Y_1 \ Y_2 \ Y_3 \ Y_4 \ Y_5 \ Y_6 ]$, where
	\begin{align*}
		Y_1 & =
		\left[
		\scalebox{0.75}{$
			\begin{array}{r}
				2 D^{3} \\
				D^{8} + 6 D^{7} + 5 D^{6} + D^{5} + 3 D^{4} + D^{3} + 4 D^{2} \\
				4 D^{13} + 6 D^{12} + D^{10} + 6 D^{9} + 3 D^{8} + 2 D^{7} + 6 D^{6} + 6 D^{5} + 4 D^{4} + 2 D^{3} + 5 D^{2} + D \\
				5 D^{13} + 2 D^{12} + 6 D^{10} + 2 D^{9} + 5 D^{8} + 4 D^{7} + 6 D^{6} + 6 D^{5} + 6 D^{4} + 4 D^{3} + 4 D^{2} + 3 D \\
				5 D^{7} + 2 D^{6} + 4 D^{5} + 5 D^{4} + D^{3} + 5 D^{2} + D \\
				3 D^{12} + 4 D^{11} + 5 D^{9} + 4 D^{8} + 3 D^{7} + D^{6} + 5 D^{5} + 5 D^{4} + 5 D^{3} + D^{2} + 3 D + 6
			\end{array}
			$}
		\right],
	\end{align*}
	
	\begin{align*}
		Y_2 & =
		\left[
		\scalebox{0.75}{$
			\begin{array}{r}
				3 D^{4} + 6 D^{2} \\
				5 D^{9} + 2 D^{8} + 2 D^{6} + 5 D^{5} + 5 D^{4} + D^{3} + 6 D^{2} \\
				6 D^{14} + 2 D^{13} + 5 D^{12} + 2 D^{11} + D^{9} + 3 D^{8} + 4 D^{7} + D^{6} + 6 D^{4} + D^{3} + 6 D^{2} + 2 D \\
				4 D^{14} + 3 D^{13} + D^{12} + D^{11} + 4 D^{10} + 3 D^{6} + 2 D^{5} + 3 D^{3} + 2 D^{2} + 2 D \\
				4 D^{8} + 3 D^{7} + 3 D^{5} + 4 D^{4} + 4 D^{3} + D^{2} + 2 D + 6 \\
				D^{13} + 6 D^{12} + 2 D^{11} + 2 D^{10} + D^{9} + 6 D^{5} + 4 D^{4} + 2 D^{2} + 4 D + 3
			\end{array}
			$}
		\right],
	\end{align*}
	
	\begin{align*}
		Y_3 & =
		\left[
		\scalebox{0.75}{$
			\begin{array}{r}
				D^{4} + 5 D^{2} + 5 D \\
				4 D^{9} + 3 D^{8} + 5 D^{7} + 4 D^{6} + 2 D^{5} + 4 D^{4} + D^{3} + D + 6 \\
				2 D^{14} + 3 D^{13} + 3 D^{12} + D^{11} + D^{10} + 3 D^{9} + D^{7} + 3 D^{6} + 5 D^{5} + 2 D^{4} + 4 D^{3} + 4 D^{2} + 3 D + 3 \\
				6 D^{14} + D^{13} + 2 D^{12} + 3 D^{11} + 4 D^{10} + 2 D^{9} + 5 D^{8} + 3 D^{7} + 5 D^{6} + 2 D^{5} + D^{4} + D^{3} + 4 D^{2} + 2 D + 1 \\
				6 D^{8} + D^{7} + 4 D^{6} + 6 D^{5} + 3 D^{4} + 6 D^{3} + 6 D^{2} + 3 \\
				5 D^{13} + 2 D^{12} + 4 D^{11} + 6 D^{10} + D^{9} + 4 D^{8} + 3 D^{7} + 6 D^{6} + 3 D^{5} + 4 D^{4} + 2 D^{3} + 3 D^{2} + D + 2
			\end{array}
			$}
		\right],
	\end{align*}
	
	\begin{align*}
		Y_4 & =
		\left[
		\scalebox{0.75}{$
			\begin{array}{r}
				5 D^{4} + 5 D^{3} + 3 D^{2} \\
				6 D^{9} + D^{7} + 3 D^{6} + 5 D^{5} + D^{4} + D^{3} + D^{2} + 3 D \\
				3 D^{14} + 4 D^{13} + D^{11} + 6 D^{10} + 4 D^{9} + 2 D^{8} + D^{7} + D^{6} + 6 D^{5} + D^{4} + 3 D^{3} + 3 D^{2} + 5 D + 1 \\
				2 D^{14} + 2 D^{12} + 4 D^{11} + 3 D^{10} + 2 D^{9} + 6 D^{8} + 6 D^{5} + 6 D^{4} + 6 D^{3} + 2 D^{2} \\
				2 D^{8} + 5 D^{6} + D^{5} + 4 D^{4} + 5 D^{3} + 3 D^{2} + 3 D + 4 \\
				4 D^{13} + 4 D^{11} + D^{10} + 6 D^{9} + 4 D^{8} + 5 D^{7} + 5 D^{4} + 5 D^{3} + 3 D^{2} + 2 D + 3
			\end{array}
			$}
		\right],
	\end{align*}
	
	\begin{align*}
		Y_5 & =
		\left[
		\scalebox{0.75}{$
			\begin{array}{r}
				5 D^{3} + 6 D^{2} + 5 D \\
				6 D^{8} + 4 D^{7} + 5 D^{6} + D^{5} + 6 D^{4} + 6 D^{3} + 2 D^{2} + 4 D \\
				3 D^{13} + 6 D^{12} + 6 D^{9} + D^{8} + 3 D^{7} + 4 D^{6} + 4 D^{5} + 3 D^{4} + 6 D^{3} + 6 D^{2} + 5 \\
				2 D^{13} + 6 D^{12} + D^{11} + 6 D^{10} + D^{9} + 4 D^{8} + 2 D^{7} + 2 D^{5} + 6 D^{4} + D^{3} + 3 D^{2} + 6 D + 1 \\
				2 D^{7} + 6 D^{6} + 4 D^{5} + 5 D^{4} + 2 D^{3} + 2 D^{2} + D + 5 \\
				4 D^{12} + 5 D^{11} + 2 D^{10} + 5 D^{9} + 2 D^{8} + D^{7} + 4 D^{6} + 4 D^{4} + 5 D^{3} + 2 D^{2} + 4 D + 4
			\end{array}
			$}
		\right],
	\end{align*}
	
	\begin{align*}
		Y_6 & =
		\left[
		\scalebox{0.75}{$
			\begin{array}{r}
				3 D^{3} + 2 D^{2} + 4 D + 1 \\
				5 D^{8} + 3 D^{7} + 5 D^{6} + 5 D^{5} + D^{4} + 4 D^{3} + 4 D^{2} + 5 D + 2 \\
				6 D^{13} + 6 D^{12} + 5 D^{10} + 6 D^{9} + 3 D^{8} + 4 D^{7} + 5 D^{5} + 3 D^{4} + D^{3} + 6 D^{2} + 3 D + 1 \\
				4 D^{13} + D^{12} + 5 D^{11} + 5 D^{10} + 3 D^{9} + 2 D^{7} + 4 D^{5} + 6 D^{4} + 2 D^{3} + 3 D^{2} + D + 5 \\
				4 D^{7} + D^{6} + 4 D^{5} + 4 D^{4} + 5 D^{3} + 6 D^{2} + 2 D + 6 \\
				D^{12} + 2 D^{11} + 3 D^{10} + 3 D^{9} + 6 D^{8} + 4 D^{6} + D^{4} + 5 D^{3} + 4 D^{2} + 2 D + 4
			\end{array}
			$}
		\right].
	\end{align*}
In this case $\Gamma(D) \neq [I_k \mid 0]$.
\end{example}



\medskip

\subsubsection{Attacks trying to factor the coefficient matrices}

Next we look at possible structural attacks to the coefficients of the generator matrix. We have
\begin{align*}
    G_0' & = S_0 G P_0, \\
    G_1' & = S_0 G P_1 + S_1 G P_0, \\
    G_2' & = S_0 G P_2 + S_1 G P_1 + S_2 G P_0, \\
    G_3' & = S_1 G P_2 + S_2 G P_1, \\
    G_4' & = S_2 G P_2.
\end{align*}

In all the cases we have constant matrices, so each of them describe a similar situation as the original McEliece PKC using block codes. Note, however, that matrices $P_i$ are neither permutations nor non singular matrices. As discussed in \cite{GR2017,karan2018,karan2019} if $G$ is a GRS generator matrix and $P$ is a matrix having columns of weight larger than or equal to 2 then no methods to factorize the product $GP$ into two matrices $\mathcal{G}$ and $\mathcal{P}$ which can be used by the attacker to decode the ciphertext are known. Due to condition \textbf{a)} in Properties \ref{conditionsT}, this is the case of the nonzero columns of the $P_i$'s. Obviously, zero columns will reveal no structure of the matrix $G$ either and thus we conclude that the attacks using the Schur square of shortenings of the public code presented in \cite{Tillich2015} will not succeed against the proposed scheme.

\medskip

\subsubsection{Structural attacks on the parity-check matrix}

Finally, we note that computing a polynomial full row rank matrix $H'(D)$ from $G'(D)$ via $H'(D)(G'(D))^T=0$ will produce a matrix of the form  $H'(D) = D^2 H (T\DD)^T $  where $H$ is a parity-check of the GRS generated by $G$. Known structural attacks to $H'(D)$ does not seem to work either as the weight of the rows of $T_{i}$ is $0$ or greater than or equal to $2$.

\section{public key sizes and ciphertext sizes}

Since we are using GRS codes, $\FFF$ and $n$ are chosen such that $n<q$ and $q$ is the greatest prime number satisfying $\lceil \log_2(q) \rceil = \lceil \log_2(n) \rceil$. The matrix $G'(D)$ is stored as the tuple of the coefficient matrices $(G'_0, G'_1, G'_2, G'_3, G'_4)$, so we need $5nk \lceil \log_2(q) \rceil$ bits to store it. On the other hand, the ciphertext $\yy(D) = \yy_0 + \yy_1 D + \cdots + \yy_{s-1} D^{s-1} \in \FFF^n[D]$ is $sn \lceil \log_2(q) \rceil$ bits long. These values are displayed in Table \ref{tabex} for different parameters of $q$, $n$, $k$ and $s$. These parameters are selected such that they provide a security level greater than of 128-bit, 256-bit and 512-bit security for the three attacks explained in Section \ref{Section ISD}. For all the proposed parameters the key size of the presented variant of the McEliece cryptosystem is significantly smaller than in other proposals, while the messages are considerably longer.

\medskip



The second part of the table exhibits parameters for some of the cryptosystems that have not been completely broken, namely,
\begin{enumerate}
	\item the classic McEliece scheme using Goppa codes;
	\item  the Niederreiter scheme based on GRS codes with a weight two mask proposed in \cite{GR2017}. In \cite[Example 3]{GR2017} public key sizes for about $80$-bit security level are proposed. In this work public key sizes with $128$-bit and $256$-bit security level against the attack in \cite{Peters2010} are proposed. The scheme in \cite{GR2017} is a variant of the BBCRS proposal \cite{BBCRS} where a weight two mask is used;
	\item the Encryption Scheme Based on expanded Reed-Solomon codes proposed in \cite{karan2019} (we exhibit the parameters from \cite[Table 1]{karan2019});
	\item the Wild McEliece schemes proposed in \cite{BLP10} and \cite{BLP11}, using wild Goppa codes with extension degree $2$. See also \cite{Tillich_IEEE2017}, where a structural attack breaking these schemes for most of the parameters proposed in \cite[Table 7.1]{BLP10} and \cite[Table 5.1]{BLP11} is described.
\end{enumerate}


\begin{table}[h!]
	\centering
	\renewcommand{\arraystretch}{1.2}
	\begin{tabular}{|l|r|r|r|c|c|c|r|r|}\hline
		&	\multicolumn{1}{c|}{$n$} &    \multicolumn{1}{c|}{$k$} & \multicolumn{1}{c|}{$s$}  &  $\WF$ \eqref{wffinal}  & $\WF$ \eqref{eq:wf_interval} & $\WF$ \eqref{wffinal1} &  Public Key & Ciphertext size\\ \hline

        &  90 & 66 & 30 & $2^{130.59}$ & $2^{129.14}$ & $2^{148.62}$ & 207900 & 18900 \\
        &  96 & 72 & 29 & $2^{131.99}$ & $2^{131.99}$ & $2^{150.42}$ & 241920 & 19488 \\
        & 108 & 72 & 24 & $2^{130.94}$ & $2^{129.47}$ & $2^{148.36}$ & 272160 & 18144 \\

        New
        & 202 & 142 & 28 & $2^{264.61}$ & $2^{257.92}$ & $2^{285.94}$ & 1147360 & 45248 \\
        Proposal
        & 220 & 148 & 25 & $2^{261.51}$ & $2^{258.31}$ & $2^{282.50}$ & 1302400 & 44000 \\
        & 244 & 160 & 22 & $2^{256.86}$ & $2^{256.86}$ & $2^{277.65}$ & 1561600 & 42944 \\

        & 396 & 288 & 29 & $2^{514.18}$ & $2^{514.18}$ & $2^{538.81}$ & 5132160 & 103356 \\
        & 408 & 288 & 28 & $2^{519.90}$ & $2^{516.23}$ & $2^{544.35}$ & 5287680 & 102816 \\
        & 420 & 300 & 28 & $2^{531.63}$ & $2^{531.63}$ & $2^{556.37}$ & 5670000 & 105840 \\

        \hline\hline

		Classic         & $2960$ & $2288$ &     & $2^{128}$   &&& $1537536$ & 672 \\
		McEliece        & $6960$ & $5413$ &     & $2^{256}$   &&& $8373911$ & 1547 \\
		& $8192$ & $6528$ &    & $2^{256}$   &&& $10862592$ & 1664 \\\hline
		GRS with        &  $784$ &  $496$ &     & $2^{128.1}$ &&& $1428480$ & 2880 \\
		Weight two mask   & $1820$ & $1384$ &     & $2^{256.0}$ &&& $6637664$ & 4360 \\  \hline
		Expanded RS     & $1258$ & $1031$ &     & $2^{256.0}$ &&& $4624198$ & 2724 \\\hline
		Wild McEliece   &  $852$ & $618$  &     & $2^{128.0}$ &&&  $\approx 712000$ & 1170 \\
		with extension  &  $858$ & $672$  &     & $2^{128.0}$ &&&  $624960$ & 930 \\
		degree $2$     &  $892$ & $712$  &     & $2^{128.0}$ &&& $634930$ & 900 \\ \hline
	\end{tabular}	
	\caption{Parameters, work factors, public key sizes and ciphertext sizes (in bits) of different PKCs}\label{tabex}
\end{table}

\section{Analysis of the computational cost complexity}

First, we analyze the number of operations on $\FFF$ necessary to encrypt a message. Since
\[ \yy(D) = \uu(D) G'(D) + \ee(D) \pmod {D^s-1}, \]
it is sufficient to compute $\uu(D)G'(D)$, take the last four blocks and add them to the first four blocks. If the message $\uu(D)$ is of length $s$ then the $i$-th coefficient of $\uu(D)G'(D)$ is given by
\[ \uu_{[i]_s} G'_0 + \uu_{[i-1]_s} G'_1 + \cdots + \uu_{[i-4]_s} G'_4, \]
so we need five vector-matrix multiplications and four vector additions. We focus on the number of multiplications as its complexity cost is higher. One vector-matrix multiplication corresponds to $kn$ multiplications between elements of $\FFF$ so the number of multiplications in $\FFF$ to encrypt a message is $5skn$. For the McEliece cryptosystem with block codes, the encryption takes approximately $(1+o(1))kn$ operations on $\mathbb{F}$. Although the proposed parameters $n$ and $k$ are considerably smaller than in the other schemes (see Table \ref{tabex}), the cost complexity is still larger due to the impact of the $s$ parameter.

\medskip

In order to decipher a message we compute first $[ \hat{\yy}_0 \ \hat{\yy}_1 \ \cdots \ \hat{\yy}_{s-1} ]$, where each $\hat{\yy}_i$ is computed as
\[
\hat{\yy}_{i} = \yy_{[i]_s} T_0 + \yy_{[i+1]_s} T_{-1} + \yy_{[i+2]_s} T_{-2}.\]
This can be done with three vector-matrix multiplications and two additions $s$ times. For the multiplication of a vector in $\FFF^n$ and a matrix in $\FFF^{n \times n}$ we need $n^2$ multiplications in $\FFF$, so in total we need $3sn^2$ multiplications in $\FFF$. To decode we need to use $s$ times any of the efficient decoding algorithms for GRS codes. For example, using Berlekamp-Massey requires $\mathcal{O}(n^2)$ operations in $\FFF$. Finally we need to solve a linear system of $sk$ unknowns and $sk$ equations. This needs $\mathcal{O}(s^3 k^3)$ operations using standard methods such as Gaussian elimination. Therefore it takes $\mathcal{O}(sn^2 + s^3 k^3)$ operations in $\FFF$ to decipher a message. For other schemes it takes around $\mathcal{O}(n^2)$ to decipher a message. Again, since the proposed parameters $n$ and $k$ are considerably smaller than in other schemes, the impact of the $s$ parameter is partly compensated, and the time it takes to decipher a message would not increase by much.

\section*{Acknowledgement}

The authors would like to thank Jorge Brandão, Karan Khathuria, Jean-Pierre Tillich and Markus Grassl for useful discussions during the preparation of this paper. The first and fourth author were supported by The Center for Research and Development in Mathematics and Applications (CIDMA) through the Portuguese Foundation for Science and Technology (FCT - Fundação para a Ciência e Tecnologia), reference UIDB/04106/2020. The second and third authors were supported by Spanish I+D+i project PID2022-142159OB-I00. 

\bibliographystyle{IEEEtran}
\bibliography{bibliography}

\end{document}